\documentclass[12pt, appendixfloats, numberedappendix]{emulateapj}
\usepackage{amsmath}
\def\eg{e.g., }

\newcommand{\mpc}{{\rm\,Mpc}}
\newcommand{\kpc}{{\rm\,kpc}}

\newcommand{\beq}{\begin{equation}}
\newcommand{\eeq}{\end{equation}}
\newcommand{\bs}[1]{\boldsymbol{#1}}
\newcommand{\norm}[1]{\left\lVert#1\right\rVert}

\def\nii{[\ion{N}{2}]}

\def\oiidoublet{[\ion{O}{2}]\,$\lambda\lambda3727,3730$}

\def\oiii{[\ion{O}{3}]}

\def\mgiidoublet{\ion{Mg}{2}\,$\lambda\lambda2796,2804$}

\def\ha{H$\alpha$}
\def\hb{H$\beta$}

\newcommand{\kms}{\ensuremath{{\rm km\,s}^{-1}}}

\usepackage{natbib}
\usepackage{rotating}
\begin{document}

\shorttitle{Set Cover Problem and Archetype}
\shortauthors{Zhu}
\title {A New View of Classification in Astronomy with the Archetype Technique: \\
An Astronomical Case of the NP-complete Set Cover Problem}

\author{
Guangtun Ben Zhu\altaffilmark{1,2}
} 
\altaffiltext{1}{Department of Physics \& Astronomy, Johns Hopkins University, 3400 N. Charles Street, Baltimore, MD 21218, USA, guangtun@jhu.edu}
\altaffiltext{2}{Hubble Fellow}

\begin{abstract}
We introduce a new generic Archetype technique for source classification and identification, 
based on the NP-complete set cover problem (SCP) in computer science and operations research (OR). 
We have developed a new heuristic SCP solver, by combining the greedy algorithm and the Lagrangian 
Relaxation (LR) approximation method.
We test the performance of our code on the test cases from Beasley's OR Library and 
show that our SCP solver can efficiently yield solutions that are on average $99\%$ optimal in terms of the cost.
We discuss how to adopt SCP for classification purposes and put forward a new Archetype technique.
We use an optical spectroscopic dataset of extragalactic sources from the Sloan Digital Sky Survey (SDSS) 
as an example to illustrate the steps of the technique. We show how the technique naturally selects a basis set of 
physically-motivated archetypal systems to represent all the extragalactic sources in the sample.
We discuss several key aspects in the technique and in any general classification scheme, 
including distance metric, dimensionality, and measurement uncertainties. 
We briefly discuss the relationships between the Archetype technique and 
other machine-learning techniques, such as the $k$-means clustering method.
Finally, our code is publicly available and the technique is generic and easy to use and expand.  
We expect that it can help maximize the potential for astrophysical sciences of the 
low-S/N spectroscopic data from future dark-energy surveys, and can find applications in many fields of astronomy, 
including the formation and evolution of a variety of astrophysical systems, such as galaxies, stars and planets.
\end{abstract}


% ==================================================================
% ==================================================================
\section {Introduction}
% ==================================================================
% ==================================================================

Astronomy is the most ancient data science and classification of celestial sources is 
one of the most ancient subjects in astronomy. The magnitude system in which
we classify stars by their apparent brightness dates back $2000$ years to Hipparchus/Ptolemy.
The modern Morgan-Keenan stellar classification system classifies stars based on 
their color, luminosity, and spectral lines \citep[\eg][]{pickering1890a, cannon1918a}. 
For (exo-)planets, we often separate them into different groups in mass/size and orbital period/radius \citep[\eg][]{borucki2010a}.
For galaxies, we usually categorize them into different morphological types, which form 
the famous Hubble tuning fork \citep[\eg][]{hubble1936a, devaucouleurs1959a}.

Classification in data science is a method in artificial intelligence, 
an extension of the tendency of human perception to rank things in order or 
put them in different groups, sometimes in a hierarchical system.
It is a powerful tool in natural science. A proper classification scheme provides a framework that 
helps us understand the underlying physics behind the appearance. 
The location of a given type of stars in the Hertzsprung-Russell diagram \citep[\eg][]{russell1914a} of the Morgan-Keenan system
reflects its physical properties, such as temperature and mass, and even its age, formation history 
and fueling mechanisms within \citep[][]{eddington1920a}. 
A giant planet (e.g., Jupiter-size) with a long orbital period is often gaseous and can be (relatively) easy 
to detect through the observation of the radial velocity of its host star \citep[the Doppler spectroscopic method, \eg][]{mayor1995a}, 
while an earth-size planet with a short period is usually rocky and it is more efficient to 
search for them with the transit photometric approach \citep[\eg][]{konacki2003a}.
Disk galaxies are composed of both old and young stars and have experienced a more extended star formation history \citep[\eg][]{larson1976a, fall1980a},
while giant elliptical galaxies host mostly old stellar populations \citep[\eg][]{trager2000a, thomas2005a, zhu2010a} 
and are believed to be remnants of violent mergers and subsequent dynamic relaxation \citep[\eg][]{toomre1977a, white1978a, naab2006a}.

Quantitative classification is usually performed in a given reduced-dimension subspace, 
or on a certain projection of the full-dimension space, and with a distance metric
that defines the similarity/distance between instances in the Universe.
One classification scheme with a specific distance metric in a given subspace
may shed light on some aspects of the instances under investigation, 
but ignoring other dimensions can also be misleading, especially if our goal is the underlying physics.
A good such example is the quasi-stellar objects (quasars). 
Quasars are point-like sources in appearance, just like ordinary stars we observe in our own Milky Way (thus the name).
They even have similar colors (spectral continuum shapes) as some types of stars, 
but spectra of these sources reveal that they are extragalactic sources at high redshift \citep[\eg][]{schmidt1963a},
with massive energy output generated by active supermassive black holes at the center of galaxies 
in the distant Universe \citep[\eg][]{salpeter1964a, zeldovich1964a, lyndenbell1969a}.
Stars and quasars, similar in the appearance, are therefore two distinctly different types of astronomical systems 
governed by completely different physical mechanisms.

Over the past two decades, the amount of data in astronomy has grown exponentially. 
This trend will only accelerate in the time to come, thanks to many large ongoing and upcoming programs,
such as 
SDSS,\footnote{\texttt{http://www.sdss.org/}}
DESI,\footnote{\texttt{http://desi.lbl.gov/}}
PFS,\footnote{\texttt{http://sumire.ipmu.jp/en/2652}}
LSST,\footnote{\texttt{http://www.lsst.org/}}
Euclid,\footnote{\texttt{http://www.euclid-ec.org/}}
and WFIRST.\footnote{\texttt{http://wfirst.gsfc.nasa.gov/}}
With the unprecedentedly large amount of data that will become available, 
one of the imminent challenges is how to efficiently extract scientific information from the unstructured data.
We expect proper classification methods, combined with other machine-learning techniques such as 
dimensionality-reduction techniques, can facilitate scientific information extraction 
and help us make better use of the data to further understand the physical Universe.

Many classification schemes can be formally formulated as quantitative (and sometimes philosophical) 
problems in computer science, applied mathematics or operations research.
Many of them are still open problems and are being actively studied in academic research.
We here discuss such a problem, the set cover problem (SCP), an NP-complete problem with no known efficient
exact algorithm. We have developed a heuristic approximation algorithm and implemented it in Python.
Adopting the problem for classification purposes, we introduce a new generic classification method,
the Archetype technique. We use an optical spectroscopic dataset of extragalactic sources from the 
SDSS survey as an example and illustrate the essence and steps of the technique.

The rest of the paper is organized as follows. In Section~\ref{sec:scp}, we introduce SCP and its formulation.
We describe the heuristic algorithm we have developed to solve SCP in Section~\ref{sec:scpsolver}. 
In Section~\ref{sec:archetype}, we introduce the new classification method, the Archetype technique and in Section~\ref{sec:discussion}, 
we discuss some key aspects of SCP and the Archetype technique and potential immediate applications of our technique.
We summarize the work in Section~\ref{sec:summary}.
When necessary, we assume the $\Lambda$CDM cosmogony, with $\Omega_\Lambda=0.7$, $\Omega_{\mathrm m}=0.3$, and ${\mathrm H}_0=70\,\kms\,\mpc^{-1}$.

% ==================================================================
\section{The Set Cover Problem}\label{sec:scp}
% ==================================================================

The set cover problem is one of the open problems in computer science and operations research. 
It has many real-life applications, such as crew-scheduling for trains and airlines, nurse scheduling,  
and location selection of facilities (e.g., fire stations and schools). 
Given a set of $m$ instances $\mathbb{M}=\left\{e_1,e_2,\ldots,e_i,\ldots,e_m\right\}$ (the universe) and a family
of $n$ subsets $\mathbb{N}=\left\{S_1,S_2,\ldots,S_j,\ldots,S_n\right\}$, whose union is the universe, 
i.e., $\cup_{S_j \in \mathbb{N}} S_j = \mathbb{M}$, 
the set cover problem is to find the subfamily (or subfamilies)
from the $n$ subsets, $\mathbb{S} \subseteq \mathbb{N}$, with the {\it minimum} cost that covers the entire universe. 
In other words, the problem is, among all the combinations of subsets from $\mathbb{N}$ whose union is the universe, 
identify the one that has the minimum cost. 
If the cost for each subset is the same, then SCP is equivalent to finding the smallest number of subsets
to include all the instances in $\mathbb{M}$. In the following formulation, 
we treat weighted set cover problem, i.e., with non-uniform costs, as the default.

% ==================================================================
\subsection{The integer linear programming formulation}\label{sec:formalation}
% ==================================================================

Mathematically, SCP can be formulated as an integer linear programming problem as follows.  
If there are $m$ rows (instances in the universe) to cover and there are $n$ columns (subsets in the family) 
to select from, we define a $m \times n$ binary matrix $\bs{A}$ to represent their relationships.
If column $j$ covers row $i$, then the element $a_{ij} = 1$, while $a_{ij} = 0$ if otherwise. 
If the cost of column $j$ is $c_j$, SCP is to find a set of columns $\mathbb{S}$ with the minimum cost $v({\rm SCP})$, 
subject to that all rows must be covered. Formally, SCP is to 

\begin{eqnarray}
\text{Minimize}\ & \bs{c}^T \bs{x} \equiv \sum\limits_{j \in N} c_j x_j \label{eq:SCP}\\
\text{subject to}\ & \nonumber \\
  & \bs{A}_i \bs{x} \equiv \sum\limits_{j \in N} a_{ij} x_j \geqslant 1\mathrm{,} \ i \in M\mathrm{,} \label{eq:allrows} \\
  & x_j \in \{0,1\}\mathrm{,}\ j \in N\mathrm{,} \label{eq:integer}
\end{eqnarray}

\noindent where we have used $M$ and $N$ to represent the set of indices of the instances in $\mathbb{M}$ 
and that of subsets in $\mathbb{N}$ for brevity. We express the solution to SCP (i.e., the minimum set) as
\begin{equation}
S=\left\{j: x_j=1\right\}\mathrm{,}
\end{equation}
\noindent and the total cost of this set as 
\begin{equation}
v({\rm SCP}) = \sum\limits_{j \in S} c_j\mathrm{.}
\end{equation}

\noindent If the costs of columns are the same ($c$), then the minimum total cost is simply $v({\rm SCP})=c\,\lvert S \rvert$, 
where $\lvert S \rvert$ is the cardinality (the number of elements) of $S$.

% - - - - - - - - - - - - - - - - - - - - - - - - - - - - - - - - - - - - - 
\begin{figure*}
%\centering
%\vspace{-2cm}
\epsscale{0.95}
\plotone{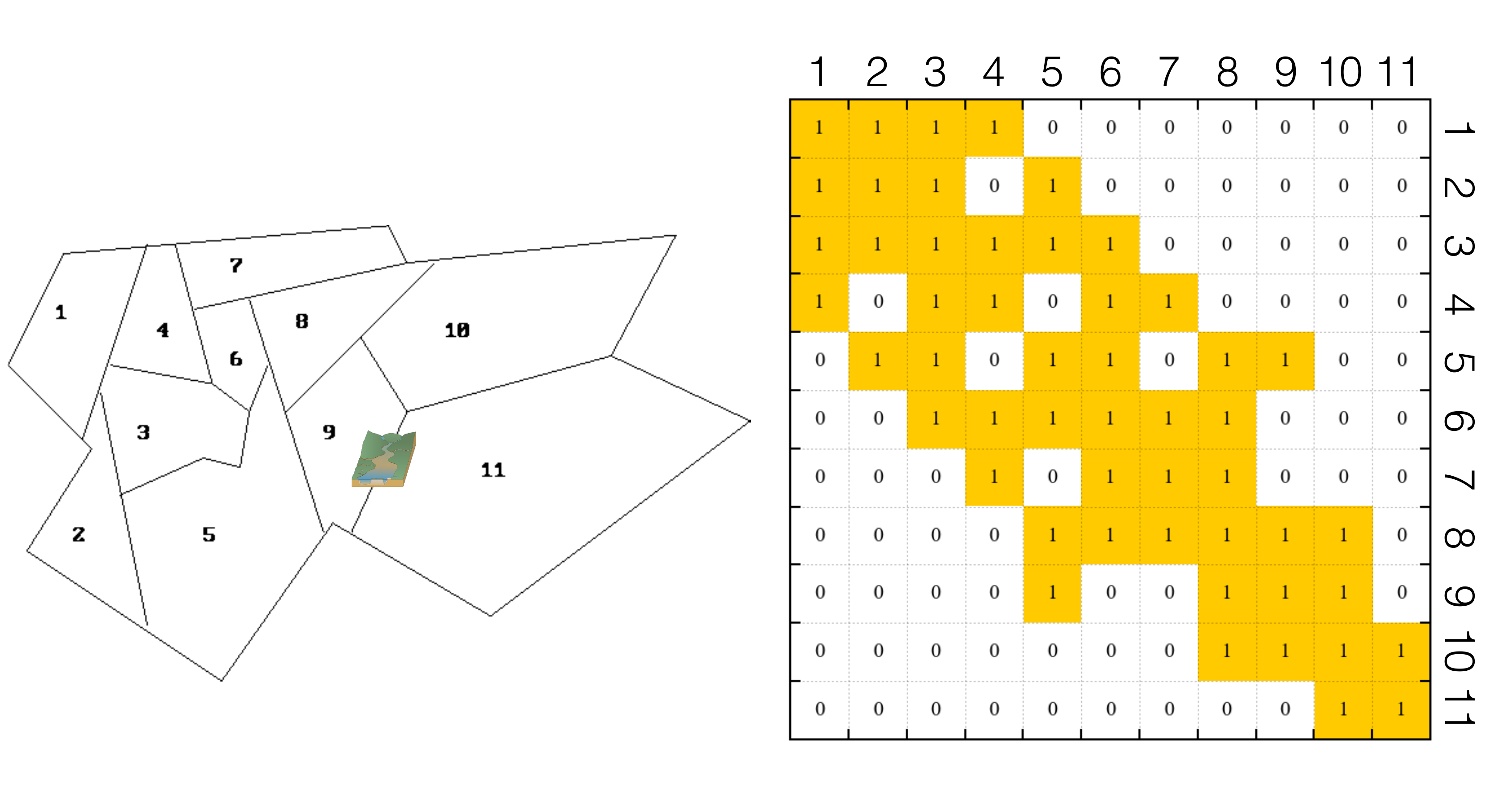}
\caption{An example of the set cover problem: build schools in a county with the minimum cost, subject to that
no child is left behind, assuming each school costs the same and students can go to the school in their home precinct 
or one in a directly neighboring precinct unless there is a natural barrier in between.
The left figure presents the geographical layout and the right figure shows the matrix form of 
the relationships between the schools and (the students in) the precincts, 
with an element value $a_{ij}=1$ (\texttt{True}) indicating that a given column (the school) $j$ 
can cover the row (precinct) $i$. There are multiple solutions to this problem, e.g., \{2,4,10\}, 
\{3,7,10\}.\footnote{Are these solutions optimal? Are there other optimal solutions? We leave these interesting questions to the reader.}
}
\vspace{0.2cm}
\label{fig:schools}
\end{figure*}
% - - - - - - - - - - - - - - - - - - - - - - - - - - - - - - - - - - - - - 

While analyzing the problem, we need to consider the reciprocal relationship between rows and columns. 
We use $I_j$ to denote the set of rows covered by column $j$, 
\begin{equation}
I_j = \left\{ i \in M: a_{ij} =1 \right\}\mathrm{,}\ j \in N\mathrm{,}
\end{equation}
and $J_i$ to represent the set of columns that cover row $i$, 
\begin{equation}
J_i = \left\{ j \in N: a_{ij} =1 \right\}\mathrm{,}\ i \in M\mathrm{.}
\end{equation}

\vspace{0.1in}
% ==================================================================
\subsection{A simple example}\label{sec:buildschool}
% ==================================================================

To understand the simplicity of the set cover problem and the complexity of its solution, 
it is instructive to consider a simple yet concrete case. 
We consider an example in which we seek to build a minimum number of schools in a county.\footnote{This example is 
a modified version of a problem from the lecture notes by Michael A. Trick, CMU (1997): 
\texttt{http://mat.gsia.cmu.edu/orclass/integer/node8.html}.}
We illustrate the problem in Figure~\ref{fig:schools}.

In this example, there are $11$ precincts in the county. If we build a school in a precinct, 
students in this precinct and its bordered precincts can attend this school, 
unless there is a natural barrier in between, such as a dangerous river. 
Under these restrictions, a school in precinct $10$ covers precincts $8$, $9$, $10$ and $11$, 
and a school in precinct $11$ covers precincts $10$ and $11$, or
$I_{10}=\left\{8,9,10,11\right\}$ and $I_{11} = \left\{10,11\right\}$ using the notations above.
We can then define a binary matrix $\bs{A}$ to describe the relationship between schools (the columns) 
and precincts (rows), which we show in the right panel of Figure~\ref{fig:schools}.
If we further assume each school costs the same,
then the problem becomes how to select a minimum number of precincts to build a school,
subject to the condition that no child is left behind.

We will refer back to this school-location example while discussing some key aspects of SCP below.
We invite the reader to think how they would solve the problem before move on to the rest of the paper. 
The caption includes more information regarding the solution. 

We would like to note that the examples we describe in this paper all have a symmetric, square 
binary relationship matrix and uniform cost for all the subsets (columns) for simplicity. 
However, the set cover problem and the algorithms we describe below do not have such restrictions.
In the school-location example, if we assume one of the precincts (say $9$) is not eligible and 
needs to be excluded and then $\bs{A}$ would become an $11\times10$ matrix, and if the costs
of schools in different precincts are different, all the discussions and methods still apply. 

% ==================================================================
\subsection{The NP completeness}\label{sec:NPcomplete}
% ==================================================================

In principle, SCP can be solved by an exhaustive search, i.e., a search over all the combinations of 
the subsets. In the school-location example, one can select one precinct, or a combination of two, or three, 
and see if any of the combinations can cover all the $m$ (11) precincts in the whole county. 
The worst-case scenario is that we need to go through all the combinations, 
in which case the total number of combinations is given by the sum of the binomial coefficients,
\begin{equation}
\sum\limits_{k=1}^{k=n} \binom{n}{k} = 2^{n}-1\,\mathrm{.}
\end{equation}
Such a brute-force approach therefore has an exponential time complexity $\mathcal{O}(m\,2^n)$. 
If we have $10^3$ instances in the dataset, the number of required operations is 
of the order of $2^{1000} \sim 10^{300}$, which is more than the total number of atoms in the whole observable Universe.
Although some techniques of data pre-processing, such as removing obvious redundant columns, can 
reduce the complexity by a small factor, there is no existing exact algorithm with polynomial time complexity.
The problem was proved to be NP-complete by \citet{karp1972a},
where NP refers to {\it non-deterministic polynomial time},
and is directly related to one of the millennium prize problems, P$\stackrel{?}{=}$NP.

A detailed discussion of the P-versus-NP problem is beyond the scope of this paper.
However, we would like to point out an important theorem, the Cook-Levin theorem \citep[][]{cook1971a, levin1973a},
which states that any problem in the NP class can be reduced in polynomial time to an NP-complete problem, 
the {\it Boolean satisfiability problem}. Based on this theorem, if one finds an efficient algorithm 
with polynomial time complexity for any of the proven NP-complete problems, such as the set cover problem,
then \textit{there exists an efficient algorithm for all the NP-complete problems, and therefore P$=$NP}, 
and vise versa, if P$=$NP, then all the NP-complete problems can be solved efficiently.

For SCP specifically, since \citet{karp1972a}, we did not find a reference that studies 
the best time complexity of an exact algorithm.\footnote{
We refer the reader to \citet[][]{woeginger2003a}, 
who conducted a survey of exact algorithms for some other NP-complete problems, such as the traveling salesman problem.}
Instead, theoretical investigations have focused on approximation algorithms \citep[\eg][]{lund1994a, feige1998a}
and tried to address what the best solution an approximation algorithm (with polynomial time complexity) can achieve.
For instance, the greedy algorithm we discuss below has time complexity $\mathcal{O}(\log n)$ and gives
an {\it approximation ratio}, the total cost of its solution divided by that of the optimal solution, 
of about $\log(n)/2$ \citep[\eg][]{johnson1974a, chvatal1979a}.
\citet{alon2006a} showed that the best approximation ratio any approximation algorithm can achieve within
polynomial time is $a\,\log(n)$, where the constant $a$ is $\lesssim0.25$.

% ==================================================================
\section{A heuristic SCP solver}\label{sec:scpsolver}
% ==================================================================

As there is no known exact algorithm to solve the set cover problem efficiently,
we resort to heuristic approximation methods, aiming at finding a (near-)optimal 
solution in a short amount of time.

The simplest heuristic approximation algorithm is the greedy algorithm, which we describe in more detail below.
Most recent effective approximation algorithms are based on {\it Linear Programming} (LP) 
relaxation \citep[][]{balinski1964a, hochbaum1982a} or {\it Lagrangian Relaxation} 
\citep[LR, \eg][]{held1970a, held1971a, geoffrion1974a}.

The main idea of LP relaxation is to relax the integer constraint on $x$ (Equation~\ref{eq:integer}), 
allowing it to be any number between $0$ and $1$: $x_j \in \left[0,1\right]$.
The relaxed LP problem can then be solved efficiently using well-known 
methods \citep[\eg][]{khachiyan1980a, karmarkar1984a}.
Starting with the optimal solution to the LP problem, one then uses heuristic methods 
such as branch-and-bound and cutting-planes to find the integer version of 
the solution \citep[\eg][]{little1963a}.\footnote{Many linear programming commercial software packages, e.g., CPLEX,
include these heuristic methods for integer linear programming problems.}

In our work, we choose to adopt the LR approach, which has provided the best existing solution 
to the standard test problems \citep[\eg][]{caprara2000a}. It is also easy to combine LR with 
the greedy algorithm and other techniques for significant improvement of the (near-)optimal solution. 
We describe the essential ingredients of our algorithms in more detail below.

% ==================================================================
\subsection{The greedy algorithm}\label{sec:greedy}
% ==================================================================

The basic idea of the greedy algorithm is as follows. We start with an empty solution set and 
select the subset (column) that covers the largest number of rows and costs the least to add to the solution. 
Then at each stage, we select the column that covers the largest number of \textit{uncovered} rows and costs the least.
We repeat the operation until the union of the solution set covers all the rows.
Let $S^*$ be the current set of columns already included in the solution set and $M^*$ be the set of uncovered rows, 
We start with $S^* = \O$ and $M^* = M$. At every stage, 
we define a score for every remaining column $j \in N\setminus S^*$ and select the one with the \textit{minimum} score 
to the solution set $S^*$. A natural choice for the score is the ratio between the cost of the column and 
the number of remaining rows covered by the column,

\begin{equation}
\sigma_j = c_j/\mu_j\mathrm{,}
\label{eq:score}
\end{equation}

\noindent where $\mu_j$ is the number of remaining rows in set $M^*$ covered by column $j$,

\begin{eqnarray}
\mu_j & = & \lvert I_j^* \rvert\mathrm{.} \nonumber \\
 & = & \lvert I_j \cap M^* \rvert\mathrm{.}
\end{eqnarray}

\noindent When combined with LR, we will modify this score definition to include the Lagrangian multiplier,
which we describe below. 

% ==================================================================
\subsection{The Lagrangian relaxation method}\label{sec:lagrangian}
% ==================================================================

The main difficulty of SCP arises from the condition that all the rows need to covered, i.e., 
the inequality constraint in the integer linear programming formulation (Equation~\ref{eq:allrows}).
The goal of the Lagrangian relaxation  method is to first relax this constraint by adding it to the 
cost function with a Lagrangian multiplier vector $\bs{u}$, 
which penalizes the violations of the constraint, and turn the original problem into an easier one.
We refer the reader to \citet{fisher2004a} for a recent review on the LR method. 
Below we describe the main ideas.

% ==================================================================
\subsubsection{The relaxed Lagrangian subproblem}\label{sec:lagrangiansubprob}
% ==================================================================

The relaxed Lagrangian subproblem reads
\begin{eqnarray}
\text{Minimize}\ & ~~\bs{c}^T \bs{x} + \bs{u}^T\,(\bs{\mathrm{I}} - \bs{A}\bs{x}) \label{eq:LagrangianSubproblem} \\
\text{subject to}\ & \nonumber \\
  & u_i \geqslant 0\mathrm{,}\ i \in M\mathrm{,}\  \\
  & x_j \in \{0,1\}\mathrm{,}\ j \in N\mathrm{,}
\end{eqnarray}
\noindent where $\bs{\mathrm{I}}$ is an identity vector with $m$ values all equal to one. 
We require the Lagrangian multiplier vector $\bs{u}$, an $m$-element vector, to be composed 
of \textit{nonnegative} values, and when the original constraint (Equation~\ref{eq:allrows}) 
is violated for a given row $i$, i.e., when $1-\bs{A}_i\bs{x} = 1$,
the cost function we want to minimize increases, 
which in turn penalizes the current solution to the Lagrangian subproblem.

The first insight why LR is an effective approximation algorithm is that 
the solution to the Lagrangian subproblem is a lower bound to the original SCP. 
To see this, assume the solution to the original SCP is $\bs{\hat{x}}$ and the solution to 
the Lagrangian subproblem is $\bs{\bar{x}}$, then
\begin{equation}
\bs{c}^T \bs{\bar{x}} + \bs{u}^T\,(\bs{\mathrm{I}} - \bs{A} \bs{\bar{x}}) \leqslant 
\bs{c}^T \bs{\hat{x}} + \bs{u}^T\,(\bs{\mathrm{I}} - \bs{A} \bs{\hat{x}}) \leqslant
\bs{c}^T \bs{\hat{x}}\,\mathrm{.}
\label{eq:LR_SCP}
\end{equation}
\noindent The first relation is true because $\bs{\bar{x}}$ is the solution to the Lagrangian subproblem,
and the second inequality is true because $\bs{\hat{x}}$ is the solution to the original SCP, which
requires $(\bs{\mathrm{I}} - \bs{A} \bs{\hat{x}}) \leqslant 0$. In reality, it is very rare that
the last two terms are equal as it requires every row is covered by exactly one column.

The second insight is that given a Lagrangian multiplier vector $\bs{u}$, 
the Lagrangian subproblem has a simple solution. 
We can simply reorganize the terms and rewrite the subproblem as
\begin{eqnarray}
\text{Minimize}\ & ~~\bs{c^T(\bs{u})}\,\bs{x} + \bs{u}^T\,\bs{\mathrm{I}}\,\mathrm{,}
\label{eq:LR_SCP_costform}
\end{eqnarray}
\noindent where the new cost vector $\bs{c(\bs{u})}$, termed the {\it Lagrangian cost} vector, is given by
\begin{equation}
\bs{c(\bs{u})} = \bs{c}-\bs{A}^T\bs{u}\,\mathrm{.}
\end{equation}
\noindent For a given Lagrangian multiplier vector, the solution to the Lagrangian subproblem is
\begin{eqnarray}
x_j = 0\ & {\rm if}\ c_j - (\bs{A^T}\bs{u})_j \geqslant 0 \mathrm{,} \nonumber \\
x_j = 1\ & {\rm if}\ c_j - (\bs{A^T}\bs{u})_j < 0 \mathrm{,} 
\label{eq:Lsub_sol}
\end{eqnarray}
\noindent since the second term $\bs{u}^T\,\bs{\mathrm{I}}$ is a constant and $x_j$ can only be $0$ or $1$.
The minimum objective function given by this solution, which we label as $L(\bs{u})$, is then a lower bound to
the original SCP.

% ==================================================================
\subsubsection{The Lagrangian Dual}\label{sec:lagrangiandual}
% ==================================================================

Since the solution to the Lagrangian subproblem, $L(\bs{u})$, is a lower bound to the original SCP 
for a given multiplier, the goal of LR is now to find the Lagrangian multiplier $\bs{u}$ that 
\textit{maximizes} $L(\bs{u})$, so that the three quantities in Equation~\ref{eq:LR_SCP} 
are (nearly) equal to each other. And this defines the Lagrangian Dual problem to the original SCP,
\begin{eqnarray}
\text{Maximize}\ & L(\bs{u}) \\
\text{subject to}\ & \nonumber \\
  & u_i \geqslant 0\mathrm{,}\ i \in M\mathrm{.}\ 
\end{eqnarray}

% ==================================================================
\subsubsection{The subgradient method}\label{sec:subgradient}
% ==================================================================

One of the popular approaches to solving the Lagrangian Dual optimization problem is the iterative method using the subgradient vector,
\begin{equation}
\bs{s} = \bs{\mathrm{I}} - \bs{A} \bs{x}\,\mathrm{,}
\label{eq:subgradient}
\end{equation}
\noindent which is a generalization of the well-known gradient descent method for differentiable cost functions.

In practice, we adopt the update rule first proposed by \citet{held1971a},
\begin{equation}
u_i^{k+1} = {\rm max}\biggl(u_i^k + \lambda \frac{{\rm UB} - L(\bs{u}^k)}{\norm{\bs{s}(\bs{u}^k)}^2} s_i(\bs{u}^k), 0\biggr),\,i\in M\mathrm{.} 
\label{eq:subgupdate}
\end{equation}
\noindent where ${\rm UB}$ is the current known upper bound, i.e., the best known solution, to the original SCP, 
$\norm{\bs{s}(\bs{u})}$ is the Euclidean ($L^2$) norm of the subgradient, and $\lambda$ is the \textit{adaptive} step size 
parameter, which can be increased or decreased depending on the rate of change in the last few iterations.

Starting with an initial guess of the solution multiplier $\bs{u}^0$, we repeat the update rule until it converges or 
a maximum number of iterations has been reached. 
We discuss how to choose the initial guess $\bs{u}^0$ in Section~\ref{sec:initialization}.

% ==================================================================
\subsubsection{New scores for the greedy algorithm}\label{sec:newscore}
% ==================================================================

Once we find the (near-)optimal solution to the Lagrangian Dual problem with the subgradient method, 
there are two ways to find the solution to the original SCP. 
One is to start with the solution $\bs{x}$ to the Lagrangian subproblem defined by the solution 
$\bs{u}$ to the Lagrangian Dual problem, and apply the greedy algorithm to the
uncovered rows, if there is any. The other is to replace the cost in the 
original SCP in the score definition (Equation~\ref{eq:score}) with the following 
Lagrangian cost \citep[\eg][]{fisher1990a} at each stage,

\begin{equation}
\gamma_j = c_j - \sum\limits_{i \in I_j^*} u_i^k\ \mathrm{,}
\end{equation}

\noindent where $I_j^*$ is the remaining (uncovered) rows covered by column $j$.  
We then apply the greedy algorithm with the following new score definition,

\begin{eqnarray}
\sigma_j = \gamma_j/\mu_j,\ & \text{if}\ \gamma_j>0\ \mathrm{,} \\
\sigma_j = \gamma_j\,\mu_j,\ & \text{if}\ \gamma_j<0\ \mathrm{.}
\end{eqnarray}

\noindent The new greedy algorithm is equivalent to solving the relaxed Lagrangian subproblem 
without the constant term ($\bs{u}^T\bs{\mathrm{I}}$ in Equation~\ref{eq:LR_SCP_costform}) for the given (near-)optimal Lagrangian multiplier $\bs{u}$,
but in addition subject to the constraint that  all the rows must be covered (Equation~\ref{eq:allrows}). 
We find this approach is particularly effective and adopt it in our solver.

% ==================================================================
\subsubsection{Iterations with new initial Lagrangian multipliers}\label{sec:initialization}
% ==================================================================

The iterative subgradient approach to the Lagrangian Dual problem may find a local instead of a global maximum.
To circumvent this issue and find a solution that is as good as possible, we can iterate all the steps with different 
initial Lagrangian multipliers $\bs{u}^0$. 

We alternate two methods to select the initial $\bs{u}^0$. In the first one, 
we generate a vector with values randomly distributed between 0 and 1. 
In the second approach, we define $\bs{u}^0$ in a greedy way, following \citet{caprara1999a},

\begin{equation}
u^0_i = {\rm min}\,\frac{c_j}{\lvert I_j\rvert},\ j \in J_i\ \mathrm{.}
\end{equation}

\noindent This choice is motivated by that columns with minimum score (low cost and many covered rows) are more likely to be in the solution,
and the rows they cover tend to have smaller multipliers to \textit{maximize} the solution to the Lagrangian Dual 
(see Equation~\ref{eq:Lsub_sol}). 
In each iteration, we also add a perturbation vector with small random values and generate a new $\bs{u}^0$ vector, 
with $u^0_i \rightarrow (1+\delta)\,u^0_i$, where the random value $\delta \in [-0.1, 0.1]$. 
This again is to maximize the chance for the subgradient optimization iterations to escape from a local maximum.

We iterate the entire procedure for a maximum 20 times or when a convergence criterion is reached.
Our experience with the cases in the test bed shows that only in a few cases could we find a marginally better solution 
with more iterations.

% - - - - - - - - - - - - - - - - - - - - - - - - - - - - - - - - - - - - - 
\begin{figure*}
\vspace{0.2cm}
\epsscale{0.85}
\plotone{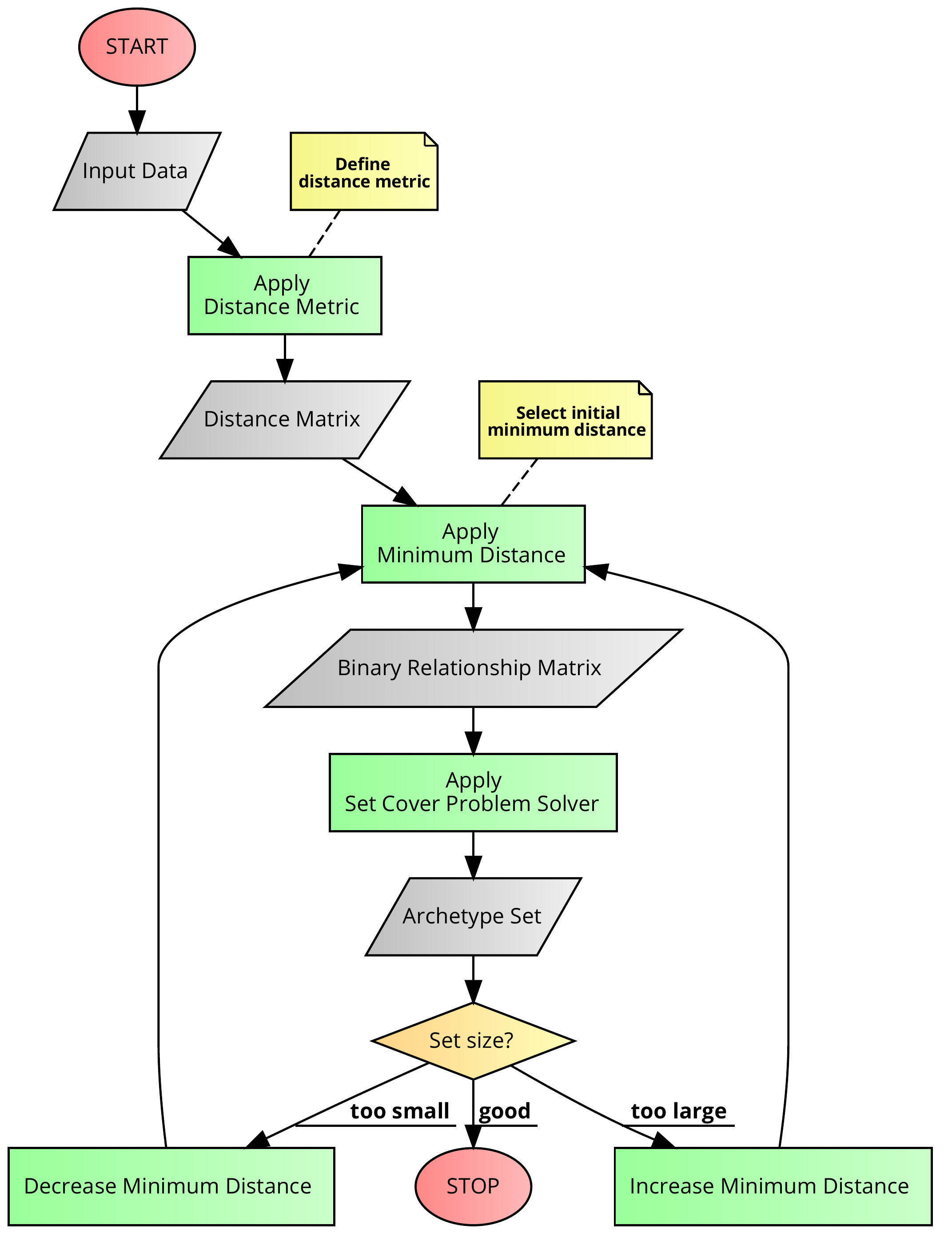}
\caption{The flow chart of the Archetype technique. Once one defines a distance metric, the minimum distance is the only free parameter. 
The final criterion, the set size, is an example for how to investigate the final basis set.
}
\vspace{0.3cm}
\label{fig:flowchart}
\end{figure*}
% - - - - - - - - - - - - - - - - - - - - - - - - - - - - - - - - - - - - - 

% ==================================================================
% ==================================================================
\subsection{The code}\label{sec:code}
% ==================================================================
% ==================================================================

We have implemented the algorithms described above in Python. 
We test our code, named SetCoverPy, on the standard test problems from Beasley's Operations Research Library \citep[][]{beasley1990a}.
We test the code on a Macbook Pro laptop with a moderate configuration of $16\,$GB RAM and $2.8\,$GHz Intel Core i7 (Quad Core).
For all the cases (4,5,6, and A-H categories), our code yields a solution that is on average $99\%$ optimal in terms of the final cost.
We provide all the test cases in convenient data format for interested readers and publish our code on the PyPI package management system. 
We briefly discuss the code and the test, and demonstrate how to install and use the code in Appendix~\ref{app:code}.

\vspace{0.1in}
% ==================================================================
\section{The Archetype technique for classification}\label{sec:archetype}
% ==================================================================

How can we adopt the set cover problem for classification purposes?
Back to the school-location example, we can think about it in a different way.
Instead of selecting a minimum number of precincts to build schools so that
every student in the whole county has a school to attend, we select a minimum number
of precincts to represent all the precincts in the whole county,
assuming that neighboring precincts are similar to each other, either in geographic
distance or by some other criteria, and they can represent each other.
Generalizing this methodology to any sample of any objects, such as animals, plants, galaxies, stars or 
planets, if we can define a distance between any pair of instances in the sample,
we can apply the same SCP solver to the data and select a minimum subset of instances, 
which we call \textit{archetypes}, that represent the whole sample. 
We introduce this generic Archetype technique for classification and describe
the key steps below. 

% ==================================================================
% ==================================================================
\subsection{The technique}\label{sec:archetypemethod}
% ==================================================================
% ==================================================================

We present the flowchart for the steps of the Archetype technique in Figure~\ref{fig:flowchart},
which we describe in detail below.
As we will demonstrate how to use the technique with a spectroscopic dataset of extragalactic sources later,
when necessary, we will assume the properties of a given instance are measured by the spectrum $f(\lambda)$, 
the flux vector as a function of wavelength. However, we would like to stress that the discussions below are 
generic and can be applied to any dataset. In other words, we can treat $\lambda$ as dimension instead of wavelength,
and $f(\lambda)$ as the location in the given dimension rather than the flux value at the wavelength.

\vspace{0.1in}
\noindent [1]. Define and apply the distance metric. The first immediate question in the technique asks 
how to measure the distance, or similarity, between a pair of instances in the dataset.
The best distance metric depends on the application, the dimensions interested, and the purpose of the distance. 
For spectral analysis, a choice often used in astrophysics is the chi-squared $\chi^2$, 
which can be considered as \textit{weighted squared Euclidean distance}. 
If we choose to scale two spectra to the same normalization with a scaling factor $a$, then the $\chi^2$ is given by 

\begin{equation}
\chi^2_{ij} = \sum\limits_{l=1}^{l=d} \frac{(f_i(\lambda_l) - af_j(\lambda_l))^2}{\sigma_{i}^2(\lambda_l) + a^2\sigma_{j}^2(\lambda_l)}\mathrm{,}
\end{equation}

\noindent and the reduced $\chi_{\rm red} ^2$ is given by $\chi^2/(d-1)$, where $d$ is the number of dimensions.
In practice, we can obtain $\chi^2$ and $a$ simultaneously by fitting the two spectra with an iterative maximum likelihood method
or a Bayesian estimator to take into account the uncertainties in both vectors \citep[\eg][]{hogg2010a, ivezic2014a}.
If we want to include the normalization (the flux level) in the metric, we can also choose to fix the scaling factor $a=1$.

A thorough discussion of distance metric is beyond the scope of this paper as 
distance metric learning itself is an active field in machine learning \citep[\eg][]{xing2003a, weinberger2009a, kulis2012a}.
We here comment on the specific usage of the (squared) Euclidean distance.
First, it is worth pointing out that using \textit{weighted} $\chi^2$ as the distance metric works the best
on a dataset with a narrow distribution of relative errors (at all dimensions). 
This is because a vector with very small errors compared to the rest of the dataset will yield very large distances to 
all other vectors by definition, while one with very large errors will yield very small distances to all other vectors,  
both of which are more likely to be selected as archetypes for the opposite 
reasons.\footnote{It therefore depends on specific applications whether to use weighted or unweighted $\chi^2$.}
Second, the definition above uses all the $d$ dimensions in the input data, in our example, all the wavelengths in the spectrum.
In the case of spectral analysis (of extragalactic sources), it is known that some wavelength regions are more informative 
about some intrinsic physical properties than the others. For example, the regions where the strong stellar absorption lines 
are particularly revealing about the stellar age and heavy element abundances in stars in the galaxy \citep[\eg][]{worthey1994a},
while those where the recombination and nebular lines are located informs mostly on the instantaneous star formation rate 
and heavy element abundances in the interstellar medium \citep[\eg][]{kennicutt1998a, kewley2008a}.

In principle, we can use a distance metric defined in any combination of the dimensions, 
or on any hyperplane (i.e., any projection), 
or in some reduced-dimension subspace (e.g., in the first few PCA component space). 
It is therefore often desirable to pre-process the data and reduce the dimensionality first. 
We can achieve this by upweighting or selecting the most informative dimensions \citep[\eg][]{yip2014a},
or projecting many correlated dimensions onto a few new dimensions defined by the most important basis components 
determined from PCA \citep[\eg][]{budavari2000a, wild2006a} or matrix factorization \citep[][]{blanton2007a, zhu2013a}, 
and defining a new distance metric with the reduced dimensionality. 
In our example, however, we will consider all the dimensions (wavelengths) provided by the observation 
for simplicity.

After defining a distance metric, we compute the distance between every pair of instances in the dataset
and obtain a (symmetric, square) distance matrix $\bs{D}$ ($\bs{\chi}^2$ in our example).

\vspace{0.1in}
\noindent [2]. Define and apply the minimum distance. To apply the SCP solver to the dataset, 
we need to select a minimum distance, within which we consider two instances are similar 
and thus can represent each other. With a chosen distance metric, this minimum distance parameter is 
the \textit{only} free parameter in the Archetype technique. 
In our example, we select a minimum chi-squared $\chi_{\rm min}^2$
and turn the $\bs{\chi}^2$ distance matrix into the binary relationship matrix $\bs{A}$:
\begin{eqnarray}
a_{ij} = 1\ ({\rm T})\ & {\rm if}\ \chi_{ij}^2 \leqslant \chi_{\rm min}^2\mathrm{,} \nonumber\\
a_{ij} = 0\ ({\rm F})\ & {\rm if}\ \chi_{ij}^2 > \chi_{\rm min}^2\mathrm{.} 
\end{eqnarray}

The freedom of choosing the minimum distance offers a degree of flexibility in the Archetype technique. 
If the minimum distance is large, then the number of archetypes will be small.
Although in this paper, for simplicity and demonstration purposes, we focus only on a given minimum distance, 
we note that varying the minimum distance can reveal, level by level, the hierarchy of the dataset.
In astrophysics in particular, it can reveal the physical mechanisms that are responsible to different degrees 
for the cosmic evolution of the astronomical systems (such as galaxies, stars and planets). 
We will discuss this further in Step 4.

A question related to specifics of the distance metric above is that whether there can be a 
forbidden connection between two instances, such as due to a natural barrier as in the school-location 
example.\footnote{I thank Adrian Liu for this interesting question.}
In the context of the distance metric (Step 1), it is not straightforward to select or add a 
dimension in which the projected distance of any pair of instances is either zero or infinity, 
because a barrier is restricted to certain pairs of instances. 
In practice, we can instead form a binary barrier matrix $\bs{B}$, in which an element is 0 (False) 
if we do not consider the corresponding pair can represent each other, 
and perform a logical AND operation between the two binary matrices.

\vspace{0.1in}
\noindent [3]. Apply the SCP solver. Once we have chosen a minimum distance and turn the distance matrix
$\bs{D}$ into the binary relationship matrix $\bs{A}$, it is now straightforward 
to apply the SCP solver (to $\bs{A}$).

One variable while applying the SCP solver is the cost to each instance. 
We here assume every instance in the dataset is equally valuable and their costs in the SCP context
are the same. In practice, it is conceivable that some instances are more valuable than the others.
However, the assignment of cost would often be \textit{ad hoc} and subjective. 
For example, considering astronomical observations, we may want to select sources with lots of
auxiliary data for in-depth investigations and therefore we may assign a smaller cost to them. 
We consider how to assign cost an open question in the Archetype technique.

\vspace{0.1in}
\noindent [4]. Investigate the basis set of archetypes and iterate the procedure. 
The flowchart shown in Figure~\ref{fig:flowchart} presents only a simple way of 
investigating the basis set by using the number of archetypes. 
In real applications, it is desirable to investigate the results more 
carefully, e.g., using pairs whose relationship is well understood 
(\textit{labeled}, as in the context of supervised machine learning).

As mentioned in Step 2, 
once we have chosen a distance metric, the only free parameter of the Archetype technique is the minimum distance
and consequently the final basis set of archetypes strongly depends on the choice.
We can view this freedom in two complementary ways. 
First, ideally we would like to select the minimum distance in an \textit{ab initio} way, 
according to some (known) strict criteria or based on our understanding of the underlying physics.
However, in most cases, we do not know if there is such an ideal minimum distance and it is often one of the goals
to find out if such a golden separation exists.
We therefore can consider in an alternative way that the freedom of choosing the minimum distance offers a degree of flexibility
and can be used to learn the different degrees of similarity among the instances and 
what different physical mechanisms are responsible at different levels. 

As common in nature, any group of objects can often be classified into a hierarchical structure.
For example, we separate plants into kingdoms, phya, classes, series, families, genera and species. 
We can iterate the procedure with different minimum distance choices and build a hierarchical classification system.
Starting with a large minimum distance, there are two options to achieve this. 
One is simply to decrease the minimum distance and apply the SCP solver to the whole dataset in each iteration, 
and the other is to decrease the minimum distance but apply the SCP solver to subsamples represented 
by each archetype in the previous iteration. 

A subtle aspect of the Archetype technique is that an instance can often be represented by more than 
one archetype, which is by design (see Figure~\ref{fig:schools}). 
When it is desirable to select \textit{the} archetype for a given instance, 
one can simply choose the one with the smallest distance.

\vspace{0.1in}
\noindent [Note]. Finally, we would like to note that, 
if computing resources are limited, we recommend to pre-process 
the initial dataset and select a subsample to choose the archetypes from. 
How to select the subsample depends on the specific application, but 
ideally the subsample should still span the whole space. 
In practice, however, our code can work on a subsample of several thousands of instances on
a typical personal computer with an average configuration (as of 2016) and yield a (near-)optimal solution 
within an hour or so (for a sample of about $3000$ instances).

% ==================================================================
\subsection{Relationships with other machine learning techniques}\label{sec:othertechnique}
% ==================================================================

The archetype technique we developed has close relationships to some of the well-known machine learning techniques,
especially in clustering analysis, such as $k$-means clustering, $k$-nearest neighbors ($k$NN) and
friends-of-friends, and in dimensionality reduction.

We first discuss a comparison of the Archetype technique with the $k$-means clustering problem. 
The $k$-means clustering problem aims at partitioning all the instances into $k$ clusters 
by minimizing the within-cluster sum of squares, i.e., sum of squared distances of each point 
in a cluster to the cluster center. The means (centers) of the clusters can also
be interpreted as archetypes, which are usually called prototypes in the context of $k$-means clustering.
It should not be surprising that the $k$-means clustering problem is also NP-hard 
and no optimal solution can be found within polynomial time \citep[\eg][]{aloise2009a}. 
The main difference between the archetype technique and the $k$-means clustering problem is 
in the free parameters, the minimum distance as opposed to the number of clusters ($k$). 
As a consequence, the clusters in $k$-means can have a wide range of scopes, 
depending on the exact way how instances are connected to each other. 
In the Archetype technique, the maximum distance within a group represented by an archetype 
cannot be larger than the minimum distance, 
while the number of archetypes depends on the overall scale of the parent sample. 
Another difference is that in the Archetype technique, an instance can be represented (covered) by
more than one archetype, while in $k$-means, as well as many other clustering/classification schemes, 
one instance only belongs to one group.

The friends-of-friends method is another popular clustering technique used in astronomy, 
especially in dark matter cosmological simulations 
\citep[][]{davis1985a}.\footnote{I thank Peter Behroozi for very useful discussions on this comparison.}
It shares the same free parameter with the Archetype technique, a minimum distance within which
two instances (e.g., dark matter particles) are considered connected to each other and belong 
in the same group (e.g., dark matter halos). However, it uses a chain connection to form
the groups: if one instance is connected to another (a friend), then it is connected to all the other 
instances connected to that friend, and all the friends connected belong to the same group (thus the name of the method).
In this regard, the friends-of-friends method, and many other distribution/connection-based clustering methods 
(such as Gaussian mixture models and $k$-means clustering),
can form groups with a wide range of scopes, as opposed to the uniform scope of groups in the Archetype technique.

As mentioned earlier, when the dimensionality is high, 
we can first pre-process the data with dimensionality-reduction
techniques, such as PCA, and define the distance metric in the reduced-dimension space 
and apply the Archetype technique. 
We refer the reader to Section~\ref{sec:archetypemethod} for a brief discussion.

% - - - - - - - - - - - - - - - - - - - - - - - - - - - - - - - - - - - - - 
\begin{figure*}
\vspace{-0.4cm}
\epsscale{1.15}
\plotone{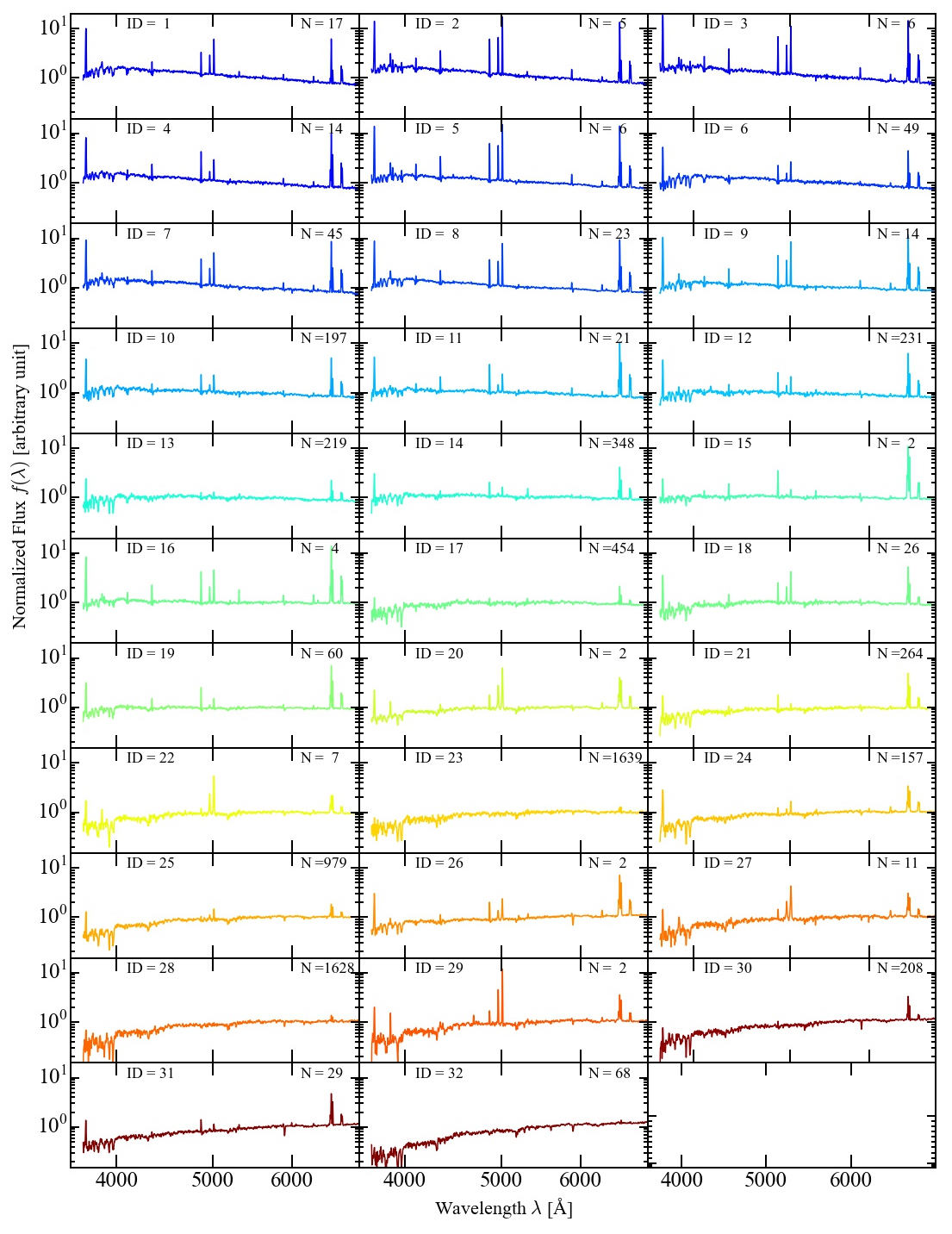}
\vspace{0.0cm}
\caption{The optical spectra of the common extragalactic source archetypes that can represent more sources than themselves, 
ordered by the continuum slope (as indicated by the color). 
The number $N$ shows how many sources in the parent dataset the archetype can represent, i.e., with distances shorter than the minimum distance.}
\label{fig:spec1}
\end{figure*}
% - - - - - - - - - - - - - - - - - - - - - - - - - - - - - - - - - - - - - 

% - - - - - - - - - - - - - - - - - - - - - - - - - - - - - - - - - - - - - 
\begin{figure*}
\vspace{-0.4cm}
\epsscale{1.20}
\plotone{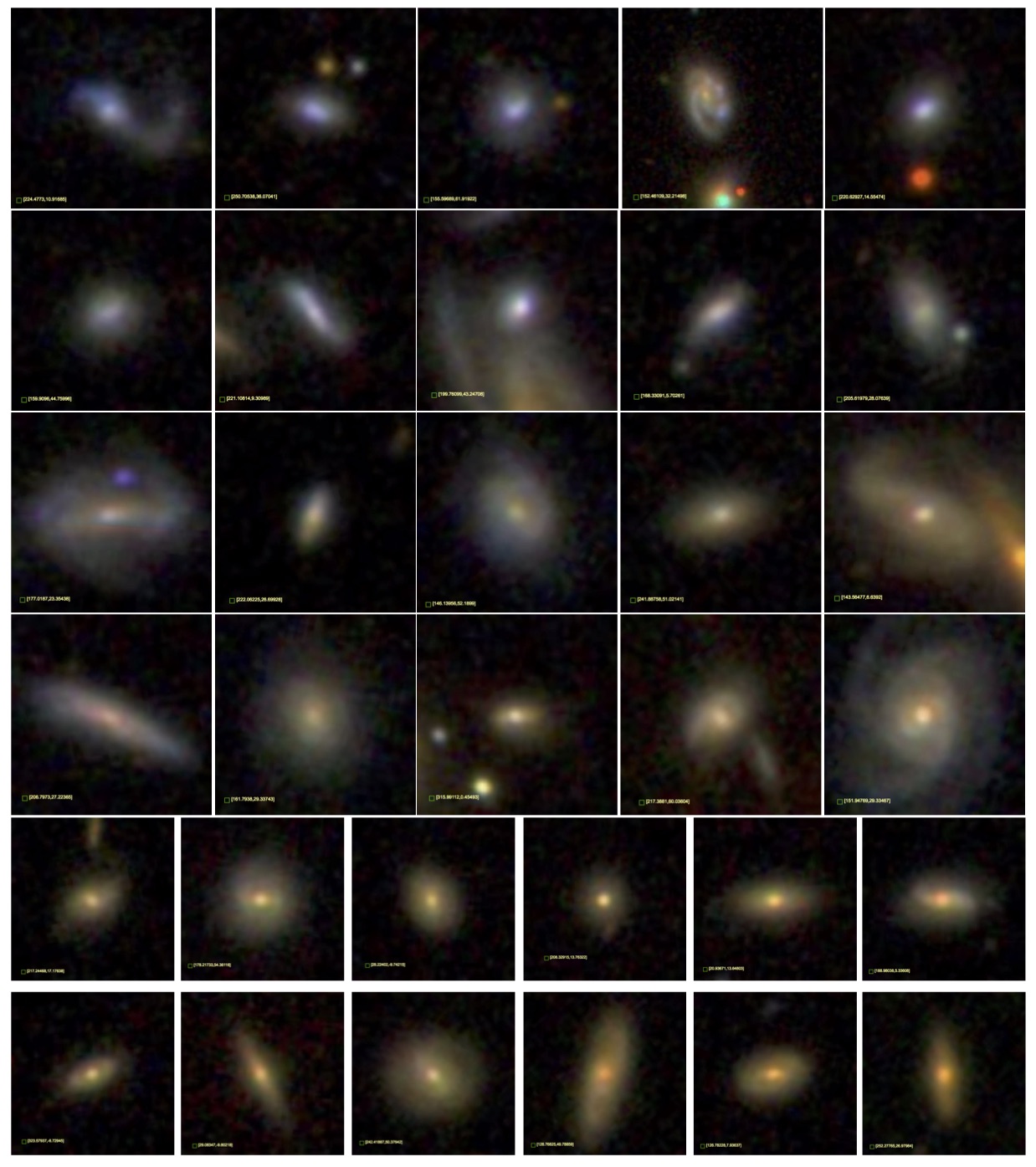}
\vspace{0.5cm}
\caption{The composite pseudo-color images of the extragalactic source archetypes that can represent more sources than themselves, ordered by the continuum slope as in Figure~\ref{fig:spec1}.
For display purposes, we have scaled down images in the last two rows to accommodate 12 archetypes.}
\label{fig:image1}
\end{figure*}
% - - - - - - - - - - - - - - - - - - - - - - - - - - - - - - - - - - - - - 

% - - - - - - - - - - - - - - - - - - - - - - - - - - - - - - - - - - - - - 
\begin{figure*}
\vspace{0.1cm}
\epsscale{1.12}
\plotone{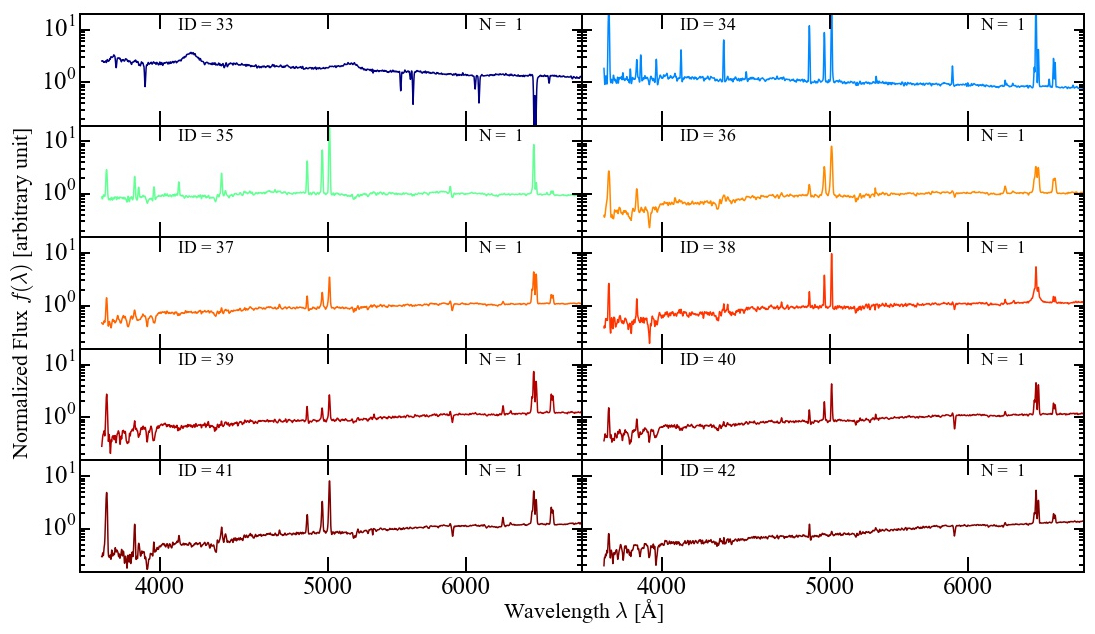}
\vspace{0.1cm}
\caption{The optical spectra of the peculiar extragalactic source archetypes that can only represent themselves,
ordered by the continuum slope (as indicated by the color).}
\label{fig:spec2}
\end{figure*}
% - - - - - - - - - - - - - - - - - - - - - - - - - - - - - - - - - - - - - 

% - - - - - - - - - - - - - - - - - - - - - - - - - - - - - - - - - - - - - 
\begin{figure*}
\vspace{0.3cm}
\epsscale{1.20}
\plotone{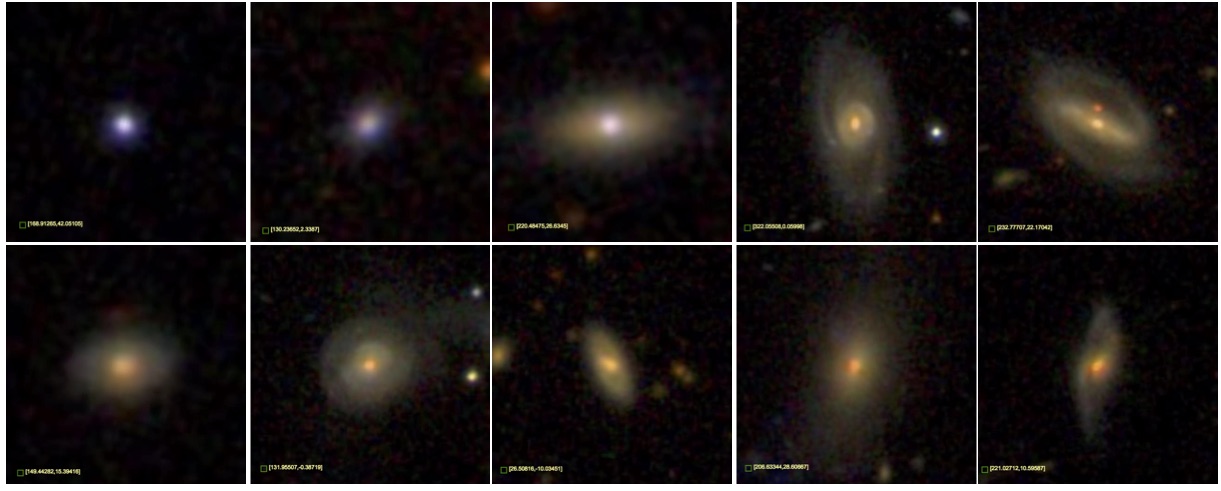}
\vspace{0.3cm}
\caption{The composite pseudo-color images of the peculiar extragalactic source archetypes that can only represent themselves, ordered by the continuum slope as in Figure~\ref{fig:spec2}.}
\vspace{0.2cm}
\label{fig:image2}
\end{figure*}
% - - - - - - - - - - - - - - - - - - - - - - - - - - - - - - - - - - - - - 

% ==================================================================
\subsection{A test case with extragalactic sources}\label{sec:archetypeexample}
% ==================================================================

% ==================================================================
\subsubsection{The parent dataset}
% ==================================================================

To further discuss the Archetype technique and illustrate how to use the method, 
we use an optical spectroscopic dataset of extragalactic sources as an example. 
We select the sources and their spectra from the seventh 
data release \citep[DR7,][]{abazajian2009a} of the SDSS legacy survey \citep{york2000a}.
In addition, we use the measurements of emission line strength and estimates of 
intrinsic properties such as stellar mass ($M^*$) and star formation rate (SFR) from the MPA-JHU 
value-added catalog \citep[\eg][]{kauffmann2003a, brinchmann2004a}.\footnote{\texttt{http://wwwmpa.mpa-garching.mpg.de/SDSS/DR7/}}

\vspace{0.1in}
We first select sources that meet the following criteria:
\begin{itemize}
\item[1.] $0.050 < z\, ({\rm redshift}) < 0.052$,
\item[2.] $15 < {\rm S/N} < 30$ and with no significant missing data.
\end{itemize}
\noindent The first selection is a compromise between the following two requirements. 
We would like to select sources at sufficiently high redshift so that the SDSS $3\arcsec$ fiber 
encompass a reasonably large area. At $z\sim0.05$, the fiber covers about $3\,\kpc$.
On the other hand, we also want to select closer systems in order to investigate 
the morphology confidently from the shallow imaging data. 

We impose the second criterion because we use the weighted $\chi^2$ as the distance metric,
and for the reasons mentioned in Step $1$ in the previous section, we try to avoid 
selecting instances with very small or large measurement uncertainties. 

Our parent test dataset includes $2820$ extragalactic sources. 
We calculate the distance matrix with the weighted $\chi^2$ by performing a least squares fitting
to every pair of spectra between $3700\,$\AA\ and $7000\,$\AA\ for the scaling factor $a$. 
Note that including the scaling factor $a$ in the $\chi^2$ calculation means we \textit{exclude} 
the normalization in the comparison, so mass or luminosity of the sources will not be 
an important dimension in our analysis.

We then choose a minimum (reduced) $\chi^2_{\rm red, min}=15$ 
to convert the distance matrix into a binary relationship matrix.
We choose this minimum distance to be concordant with the S/N selection criterion.
However, we would like to remind the reader that the minimum distance is a free parameter in the technique 
and varying it can help construct a hierarchical classification system and 
reveal different degrees of similarity among the instances and 
physical processes that are responsible at different levels.
As our goal is to describe the key steps involved in the technique, 
we have selected this particular minimum distance for the convenience of presentation.

Finally, we treat all the sources equally and assign equal cost. 
Applying the SCP solver to the dataset, we establish a basis set of $42$ extragalactic source archetypes. 
We investigate these archetypes and their relationships with the instances in the parent sample below.

% - - - - - - - - - - - - - - - - - - - - - - - - - - - - - - - - - - - - - 
\begin{figure*}
\vspace{0.1cm}
\epsscale{0.36}
\plotone{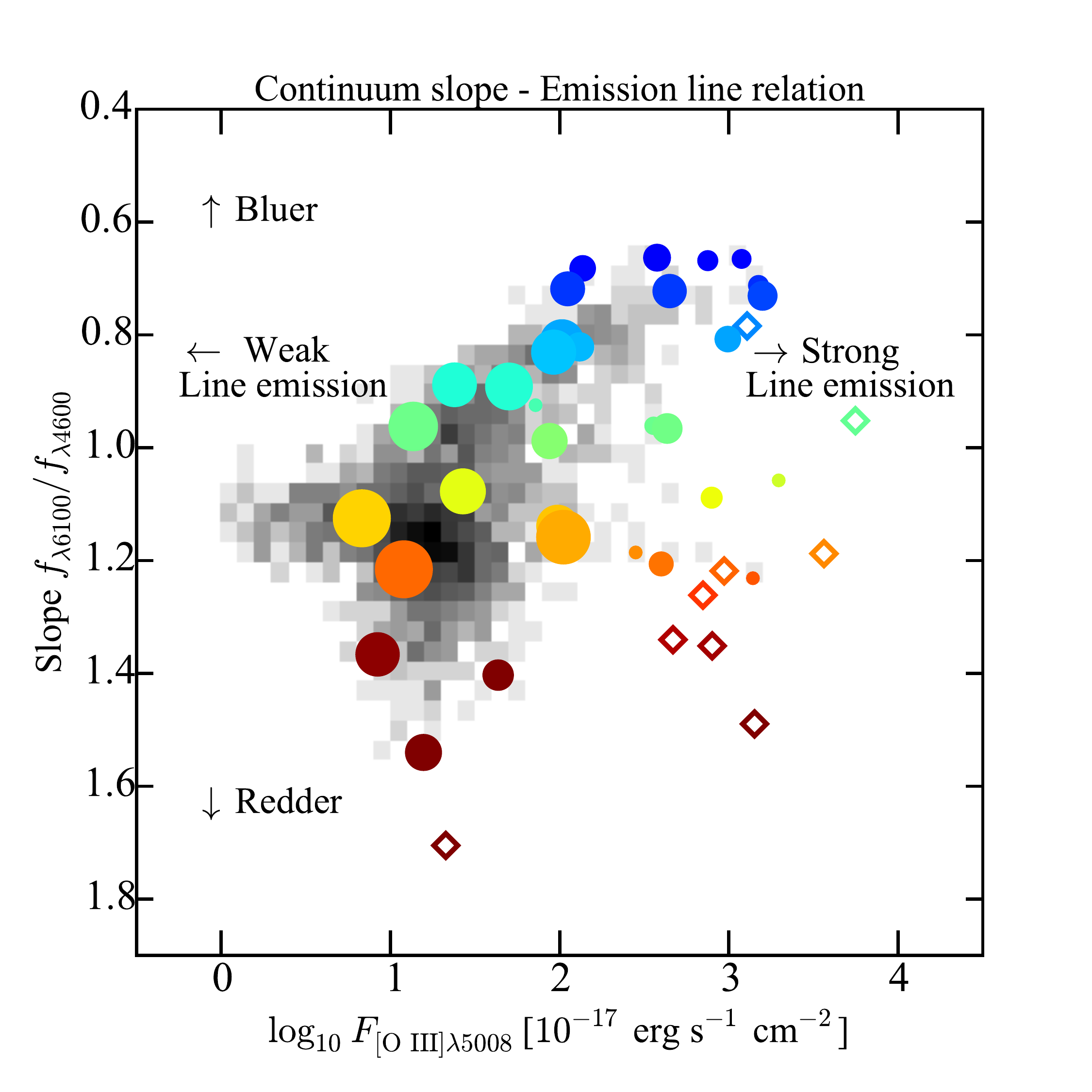}
\epsscale{0.36}
\plotone{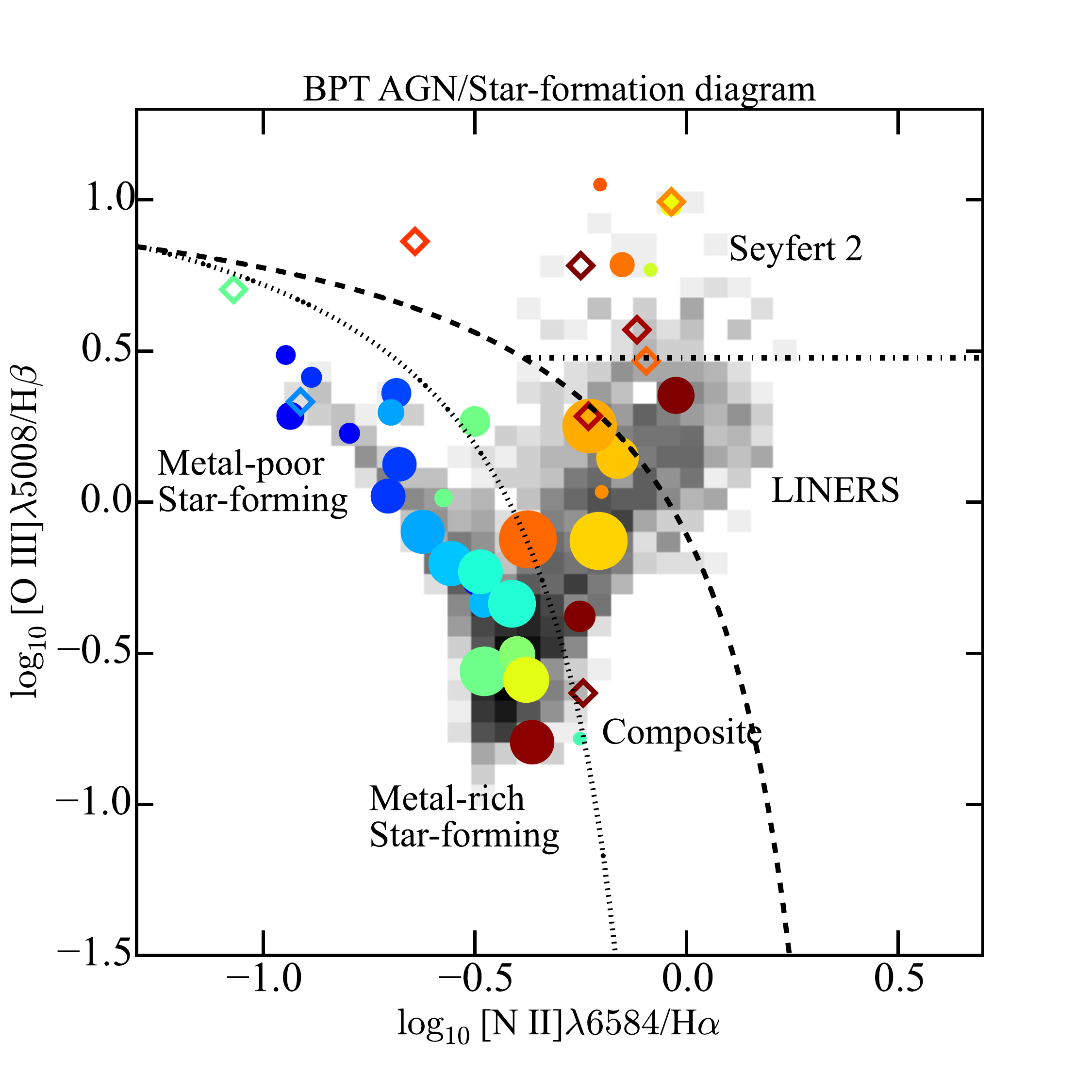}
\epsscale{0.36}
\plotone{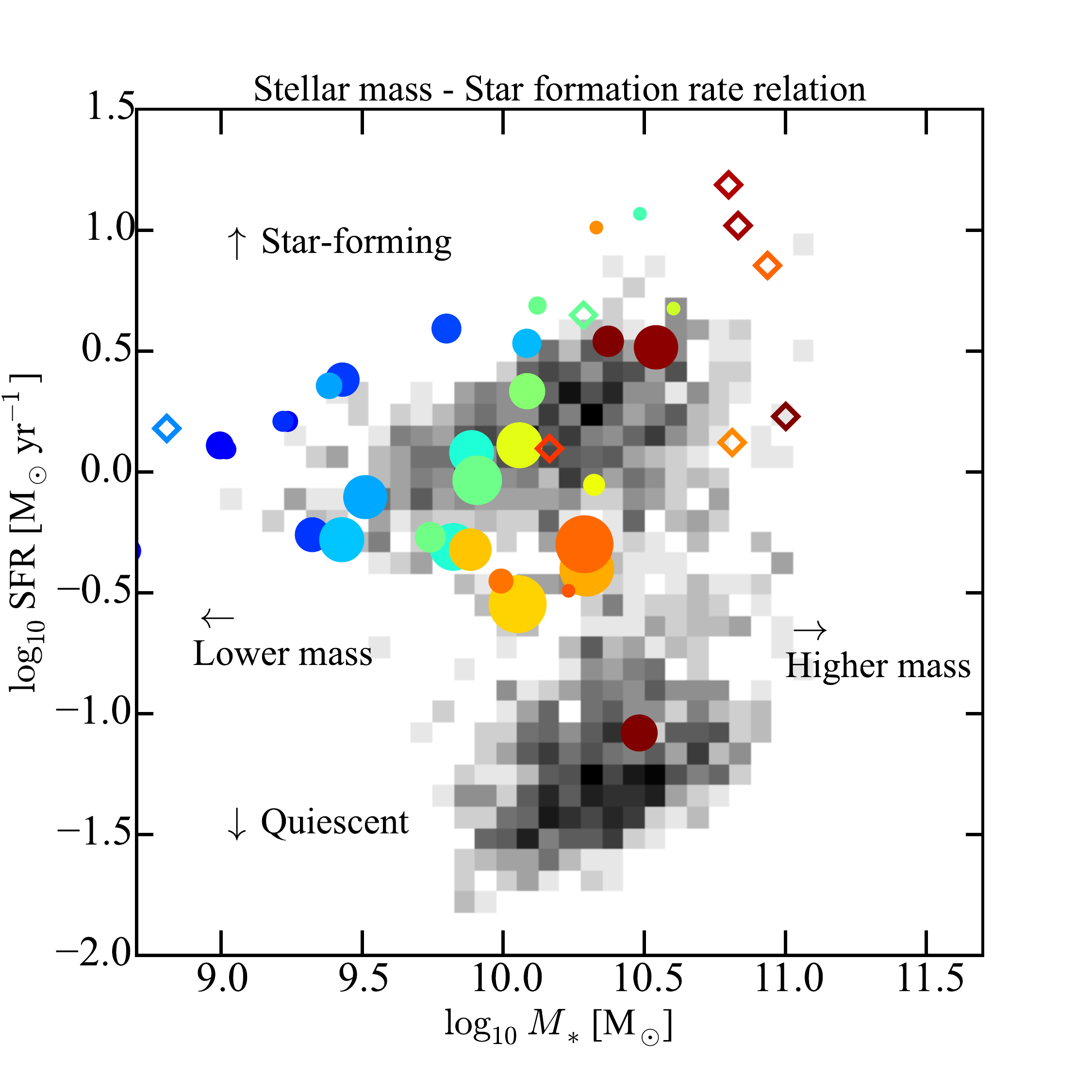}
\vspace{0.1cm}
\caption{The basis set of archetypes for extragalactic sources. 
The color indicates the continuum slope and the symbol size shows how many other sources the archetype can represent, 
with the open diamonds showing separately peculiar archetypes that can only represent themselves. 
The gray scales show the density distribution of the parent dataset.
\textit{Left}: the distribution of the continuum slope and the \oiii$\,\lambda5008$ line flux. 
\textit{Middle}: the BPT classification diagram for AGN and star-forming galaxies.
\textit{Right}: the distribution of two derived parameters, stellar mass ($M^*$) and star formation rate (SFR). 
Note we included a scale (normalization) factor in the $\chi^2$ distance metric, 
so mass is not an important dimension in this particular classification scheme.
Most of the quiescent galaxies are represented by the three archetypes (23, 25, 28) with slope $\sim1.2$ for the particular $\chi^2_{\rm min}$ choice
in this example.
}
\vspace{0.6cm}
\label{fig:GalaxyArchetypeProp}
\end{figure*}
% - - - - - - - - - - - - - - - - - - - - - - - - - - - - - - - - - - - - - 

% ==================================================================
\subsubsection{The basis set: SED and morphology}
% ==================================================================

To investigate the basis set of the archetypes, we divide the archetypes into two subsets.
One subset includes common archetypes that can represent more than themselves in the parent sample,
and the other includes peculiar archetypes that can only represent themselves.

Figure~\ref{fig:spec1} shows the spectra of the subset of common archetypes, 
in the order of the slope of the underlying continuum, the ratio between the fluxes 
at $6100\,$\AA\ and $4600\,$\AA\ ($f_{\lambda6100}/f_{\lambda4600}$). 
We choose the two wavelengths to be where the continuum is relatively smooth. 
For normal galaxies, a bluer continuum (more flux at shorter wavelength) roughly means the 
there are more younger stars in the galaxy.
Note we have normalized the spectra since we do not consider the overall brightness.
These archetypes span a variety of spectral types, as indicated by the diverse array of continuum shapes and 
the emission line strengths and ratios. An interesting observation is that the three archetypes that cover most
instances, with ID $23$, $25$, and $28$ as shown in the figure, are moderately red spectra that show little line emission.
This means the emission lines are a major source of the spectral variety and must be responsible for 
a large fraction of the (high) dimensionality in the spectral space.

Figure~\ref{fig:image1} presents the pseudo-color composite images of these archetypes from $g$-, $r$-, $i$-band 
imaging data.\footnote{Retrieved from \texttt{http://skyserver.sdss.org/}.}
Unsurprisingly, they display different morphologies, including irregular, spiral, lenticular and elliptical shapes.

We now take a look at the subset of peculiar archetypes that can only represent themselves 
in Figure~\ref{fig:spec2} and \ref{fig:image2}, again ordered by the continuum slope. 
The first one turns out to be an error of the observation.
It is actually a high-redshift quasar with an strong absorption system induced by 
a foreground gaseous cloud.\footnote{For interested readers, the SDSS Plate-MJD-Fiber of this object 
is 352-51694-380 and its equatorial coordinates are (${\rm RA}=258.20837\,{\rm deg}$, ${\rm DEC}=64.05295\,{\rm deg}$).
To learn more about intervening absorption-line systems, see \citet{zhu2013a} and references therein.} 
The reason it was identified as a $z\sim0.05$ galaxy is because one of the absorption lines, the \mgiidoublet\ doublet,
is treated as (negative) \ha\ emission by the SDSS reduction pipeline. The pipeline fit the spectra with 
linear combinations of PCA components, without nonnegativity constraint. It is interesting to see that the Archetype technique
can automatically pick out such misidentifications. We exclude this source in the further discussion below.

The second and third peculiar archetypes (with ID $34$ and $35$) exhibit extremely strong emission lines, 
indicating extreme star-bursting behaviors. 
Other peculiar archetypes share a common trait that they tend to have a red continuum but also strong and often broad emission lines, 
which indicates the presence of an AGN in an otherwise quiescent galaxy. 
Their colors and morphologies shown in Figure~\ref{fig:image2} also support this proposition. 
The main differences among them are the different line strengths and ratios.

% ==================================================================
\subsubsection{The basis set: the distribution in the reduced-dimension space}
% ==================================================================

To further investigate the basis set of archetypes, we compare them with the parent sample 
in the reduced-dimension space. We do not perform complex dimensionality reduction, 
but rather define/pick the dimensions manually according to our understanding of the spectral energy distribution (SED) and the underlying physics.
Figure~\ref{fig:GalaxyArchetypeProp} shows the comparison, where the color indicates the continuum slope 
and the symbol size represents the number of instances the archetype can represent, with the open diamonds separately 
showing the peculiar archetypes that can only represent themselves. 
The gray scale shows the distribution of the parent sample.
In the left panel, we plot the distribution of the continuum slope and the \oiii$\,\lambda5008$ emission line strength. 
These are likely the two of the most informative dimensions in our analysis, i.e., they affect the $\chi^2$ value the most.
The archetypes that can represent the largest number of sources are located at the densest regions of this diagram, 
while peculiar archetypes and those that can only represent a small number of sources have strongest emission lines,
suggesting a wide range of emission line strengths/ratios are responsible for driving the dissimilarity 
(larger $\chi^2$ distance) between them and the others. 

What are the blue archetypes with strong emission lines that can only represent a small sample of similar sources? 
We can find the answer in the middle panel, where we present the AGN-star-forming galaxy diagnostic 
diagram, the so-called BPT diagram \citep{baldwin1981a}. 
The BPT diagram compares the ratio of \oiii$\,\lambda5008$ to \hb\ and 
that of \nii$\,\lambda6584$ to \ha\ and can efficiently distinguish different types of 
extragalactic sources with different physical properties. 
We have overplotted the demarcation lines for star-forming galaxies empirically-defined 
by \citet[][dotted line]{kauffmann2003a} and theoretically-defined by \citet[][dashed line]{kewley2001a}, 
and a horizontal line at \oiii$\,\lambda5008$/\hb$=0.3$, the conventional criterion to separate 
Seyfert 2 galaxies \citep[][]{seyfert1943a} and low-ionization nuclear emission 
regions \citep[LINERs,][]{heckman1980a}.\footnote{The nature of LINERs is still under debate.}
Looking at the blue archetypes first, we see they basically follow the left star-forming sequence. 
The blue archetypes that can only represent a few sources are mostly metal-poor star-forming galaxies 
with large \oiii$\,\lambda5008$-to-\hb\ line ratios. For the peculiar archetypes with red continuum slope but strong 
and often broad emission lines, they are located in the upper right corner, 
identified as Seyfert 2 galaxies, galaxies that host Type 2 AGN at the center. 
The common archetypes with weak emission that can represent many sources are mostly distributed
in between, representing metal-rich star-forming galaxies, composite (with both moderate star formation and AGN activities), 
and LINERs.

The right panel shows two \textit{derived} properties of the extragalactic sources, 
the stellar mass ($M^*$) and instantaneous star formation rate (SFR). 
We remind the reader that we fit for the scaling factor when comparing
every pair of sources, thus the stellar mass, roughly proportional to the luminosity, 
is not an important dimension.
The metal-poor star-forming galaxies tend to be low-mass systems that are undergoing star-bursts (strong star formation events).
The peculiar archetypes are mostly massive galaxies.
However, their SFRs are likely severely over-estimated due to AGN contribution to the emission lines. 
Their correct positions in this diagram are likely an order-of-magnitude down.
With our minimum distance choice, the quiescent galaxies can be represented by the few archetypes with 
little line emission (${\rm ID} = 23$, $25$ and $28$). 
We note that SFRs (for relatively quiescent galaxies) derived from weak line emission are uncertain 
and should not be taken at the face value.
In order to select a basis set of archetypes that can properly represent quiescent galaxies in this space, 
one also needs to include the luminosity as a factor in the distance metric, i.e., use the absolute flux 
without the scaling factor $a$, and also use more robust estimates of the intrinsic properties.

To summarize, we have demonstrated that the Archetype technique we developed can naturally select a basis set 
of physically-motivated archetypes to represent the whole universe of extragalactic sources. 
In the next section, we discuss further the potential applications of this new technique.

% ==================================================================
\section{Further Discussions}\label{sec:discussion}
% ==================================================================

We have introduced a new generic classification technique, the Archetype technique, by adopting the set cover problem. 
As we discussed earlier in Section~\ref{sec:othertechnique}, there are many different ways of categorizing a sample of objects.
Nature is often more complex than what Occam's razor suggests, and there is not a golden method that works for everything.
For different purposes, it is therefore often necessary to use different approaches or certain combinations of them.
We briefly discuss some of the astrophysical applications in which the Archetype technique can be particularly useful.

[1]. Classification and identification. 
The obvious application is classification and identification of any astronomical sources: galaxies, stars or planets. 
As we discussed earlier, we can also vary the minimum distance and/or combine with other techniques, 
such as $k$-means or PCA, to study different aspects of the data or build a hierarchical system. 
For example, in the test case with the extragalactic sources, we can separate the sources into 
star-forming galaxies and non-star-forming systems, and then divide star-forming galaxies into 
metal-poor and metal-rich sub-populations. We can then study the intrinsic properties and different physical mechanisms 
responsible for the different types of systems and their cosmic formation history.

[2]. Redshift determination in future dark-energy surveys. 
Future dark-energy surveys, including SDSS-IV/eBOSS \citep{dawson2016a},
DESI \citep{schlegel2011a, levi2013a}, and PFS \citep{takada2014a}, 
aim to obtain optical spectra in the observer frame for tens of millions of galaxies 
at redshift $0.6\lesssim z \lesssim 2.0$ and measure the scale of the baryon acoustic oscillation \citep[BAO, \eg][]{cole2005a, eisenstein2005a}
as a function of cosmic time and chronicle the expansion history of the Universe. 
A primary large-scale structure tracer these surveys will target is 
the emission-line galaxies \citep[\eg][]{zhu2009a, comparat2016a}, 
for which the redshift can in principle be well-determined by their strong emission lines.
For DESI and PFS, the line of interest is \oiidoublet\ and spectroscopy with 
resolution $\mathcal{R}\gtrsim4000$ can tell apart the two components and identify the doublet.
For eBOSS, the resolution is lower, but because the targets are at lower redshift, 
other lines at longer wavelength can be used to help measure the redshift.
All these surveys are only interested in the redshift and pushing the exposure time 
to the shortest limit in order to maximize the survey efficiency and minimize the cost.
The low S/N of the spectroscopy makes even the determination of the redshifts 
difficult, especially due to the strong telluric emission lines (e.g., hydroxyl lines) 
in the background that can mimic the emission lines.

We have learned that the emission lines strengths and ratios span a wide range and are 
some of the most important dimensions in the spectral space (see the above section). 
Redshift surveys usually fit the observed spectra with linear combinations of some
basis templates, such as components from PCA or matrix factorization analysis, 
that are shifted to different redshifts, 
and find the redshift and the combination that give the least $\chi^2$.
The low S/N of the spectroscopy and the contamination from telluric lines, 
however, means it is very easy to find a good fit to the data with a wrong combination of the templates. 
In other words, the method could find a wrong answer by overfitting the low-S/N data 
with a completely unrealistic model that does not even exist in the real Universe. 

One of the advantages of the Archetype technique is that the archetypes 
would be all real systems that actually exist in the Universe.
If we fit the observations with realistic archetypal spectra, then we do not need to 
worry that we would be overfitting the data with unrealistic instances in the unoccupied space.
An earlier example of using archetypes for redshift determination is the deep low-resolution prism 
spectroscopic PRIMUS survey \citep{coil2011a}.
To overcome the challenge caused by the low resolution and low S/N of their prism spectra, 
they selected archetypes at low redshift from the AGES survey \citep[\eg][]{moustakas2011a} with the CPLEX package, 
manually modified the basis set based on the understanding of the evolution of the emission lines and shapes 
of the galaxy SEDs, and used the modified set of spectra to fit the observed spectra. 
Compared to using linear combinations of templates from PCA or matrix factorization analysis, 
this approach has helped improve the success rate of redshift determination \citep{zhu2011thesis, cool2013a}.

For dark-energy surveys, we expect to apply the Archetype technique iteratively to the redshift determination.
We start from the sample of objects with robust redshift measurements and 
select a first basis set of archetypes. We then iterate the archetype set construction and redshift determination,
each time with a better and more complete set of archetypes, until we reach the maximum success rate. 
At each stage, instead of using the low S/N spectrum of a given archetype, 
we can construct a high S/N composite spectrum of the sources that can be represented by this archetype 
and use this composite as the new archetype.
We are currently investigating this new method in the eBOSS survey,
and we expect that it will also help maximize the efficiency of future dark-energy surveys.

\vspace{0.1in}
[3]. Astrophysical sciences with low-S/N data. 
The large amount of data provided by the ongoing and future surveys offers a good
opportunity for making unexpected discoveries. One of the major challenges to extract
the scientific information is the low S/N of the observation of any individual source.
We will have to resort to composite analysis in order to enhance the S/N by orders-of-magnitude for any robust measurement. 
For example, \citet{zhu2015a} stacked the \textit{entire} dataset of about $9000$ spectra 
from the pilot observation of emission-line galaxies in eBOSS in order to 
robustly measure the resonant absorption and nonresonant fluorescent emission in the near-ultraviolet,
the signatures of galactic-scale outflows associated with star formation.
While statistical stacking has its merits, it also has some major caveats 
as one may be averaging over instances that are distinctly different systems.\footnote{As the old joke says: 
how many legs do you get if you average over humans and dogs?}

We can use the Archetype technique to mitigate the selection bias 
in composite analysis and construct more sensible subsets of same type of sources for further investigations.
More specifically, we can select the most informative dimensions for the physics we are 
interested in (with the previous knowledge as a prior),
define a distance metric in the reduced-dimension space, build a basis set of archetypes, and then perform
composite analysis for each group of sources that can be represented by a given archetype. 
Again, take the high-redshift emission-line galaxies from the eBOSS survey as an example, 
since we have learned that the emission lines are some of the most important dimensions (e.g., see the above section), 
we can select the wavelength regions where the lines are located and then apply the Archetype technique. 
In a sense, stacking the observations of sources represented by a given archetype 
is equivalent to taking many exposures (or a very long exposure) of a single source.
We can then use the composite observations to study the underlying physical mechanisms 
and the formation histories of different types of sources (Zhu et al. in prep).
We expect that this new way of analyzing the big data will be particularly useful for 
the low S/N data from future dark-energy surveys.

\vspace{0.1in}

% ==================================================================
\section{Summary}\label{sec:summary}
% ==================================================================

Astronomy is a data science. As in many other fields, the amount of data has grown drastically in astronomy 
and will continue to increase exponentially thanks to all the ongoing and upcoming large programs.
How to efficiently extract scientific information from the unstructured data 
has now become one of the major challenges in the field.
One of the important and interesting problems is to classify and identify sources 
into families or types, ideally based on intrinsic properties.
A proper classification scheme can further the understanding of the roles of different physical processes 
that govern the formation and evolution of different types of astronomical sources.
We have introduced a novel classification method, the Archetype technique, 
based on the NP-complete set cover problem (SCP) in computer science and operations research. 

We first introduced SCP and in particular discussed the simplicity of
the problem and the complexity of its solution, the NP-completeness.
We have developed a heuristic solver in Python, 
by combining the greedy algorithm and the Lagrangian Relaxation approximation approach.
We have tested the performance of our code on the standard test cases from Beasley's OR Library and 
shown that our code can efficiently produce solutions that are on average $99\%$ optimal. 

Adopting SCP for classification purposes, we introduced the Archetype technique. 
Based on how similar the sources are to each other, the Archetype technique finds a basis set of
archetypes to represent the whole universe of the sources. 
We described the steps of the technique, paying special attention to the distance metric and 
the only free parameter, the minimum distance within which two sources can represent each other.

We used a spectroscopic sample of extragalactic sources from the SDSS survey as 
an example to illustrate how to apply the technique and how to interpret the results. 
We showed that the technique naturally selected a basis set of 
physically-motivated archetypes for the extragalactic sources. 
The archetypes include different types of sources, such as metal-poor/rich star-forming galaxies, 
AGN and composite systems, and span a wide range in the spectral energy distribution and morphology. 
We show that the line emission strengths and ratios are important dimensions in the spectral energy distribution
and suggest that dark-energy surveys targeting emission-line galaxies can use the Archetype
technique to improve the survey efficiency.

We further discuss the potential future applications of our technique.
We discuss how to apply it to the low-S/N spectroscopic data and maximize the potential 
for astrophysical sciences of future dark-energy surveys, 
Our technique is generic and is easy to use and expand, and we expect that it can find applications in many fields of astronomy, 
including the formation and evolution of a variety of astrophysical systems, 
such as galaxies, stars and planets.

\acknowledgments

G.B.Z. wishes to thank Sam Roweis and David Hogg for introducing the set cover problem to him and
for very illuminating discussions. 
He also thanks John Moustakas for very useful discussions and comments that helped improve the clarity of the paper. 
He would also like to thank Mike Blanton for very useful discussions.
G.B.Z. acknowledges support provided by NASA through Hubble Fellowship grant \#HST-HF2-51351 awarded by the Space Telescope Science Institute, which is operated by the Association of Universities for Research in Astronomy, Inc., under contract NAS 5-26555. 

This paper uses the public data from the SDSS survey. Funding for the SDSS and SDSS-II has been provided by the Alfred P. Sloan Foundation, the Participating Institutions, the National Science Foundation, the U.S. Department of Energy, the National Aeronautics and Space Administration, the Japanese Monbukagakusho, the Max Planck Society, and the Higher Education Funding Council for England. The SDSS Web Site is http://www.sdss.org/.

The SDSS is managed by the Astrophysical Research Consortium for the Participating Institutions. The Participating Institutions are the American Museum of Natural History, Astrophysical Institute Potsdam, University of Basel, University of Cambridge, Case Western Reserve University, University of Chicago, Drexel University, Fermilab, the Institute for Advanced Study, the Japan Participation Group, Johns Hopkins University, the Joint Institute for Nuclear Astrophysics, the Kavli Institute for Particle Astrophysics and Cosmology, the Korean Scientist Group, the Chinese Academy of Sciences (LAMOST), Los Alamos National Laboratory, the Max-Planck-Institute for Astronomy (MPIA), the Max-Planck-Institute for Astrophysics (MPA), New Mexico State University, Ohio State University, University of Pittsburgh, University of Portsmouth, Princeton University, the United States Naval Observatory, and the University of Washington.

\appendix

% ==================================================================
\section{The code}\label{app:code}
% ==================================================================

We have implemented the algorithms for the set cover problem described in Section~\ref{sec:scp} in Python 3.
We have made extensive use of NumPy\footnote{\texttt{http://www.numpy.org/}} and 
the \texttt{sparse} matrix package from SciPy\footnote{\texttt{http://docs.scipy.org/doc/scipy/reference/sparse.html}} 
for high-performance set operations.

% ==================================================================
\subsection{Install the package}
% ==================================================================

We share our code, named \texttt{SetCoverPy}, on the repository hosting service GitHub and 
interested user can fork or clone the repository.\footnote{\texttt{https://github.com/guangtunbenzhu/SetCoverPy}}
For readers who are only interested in using the package, we have also published the package on Python Package Index (PyPI) 
and one can install it with:

\begin{verbatim}
  > pip install SetCoverPy
\end{verbatim}

We recommend the latter approach unless the user is particularly interested in helping further development of the code,
as the indexed version has been more properly tested. 
The repository webpage also includes documentation and a brief user guide of the package, which will be maintained 
and updated by the author.

% ==================================================================
\subsection{Beasley's OR Library}
% ==================================================================

To test the performance of our code, we have compiled the $65$ test problems in the 4, 5, 6 and A-H categories 
from Beasley's OR Library \citep[][]{beasley1990a}.\footnote{\texttt{http://people.brunel.ac.uk/\%7emastjjb/jeb/info.html}} 
For convenience, we have converted all the instances into NumPy data format 
(.npy), which we distribute to the public,\footnote{\texttt{http://www.pha.jhu.edu/\%7egz323/scp/BeasleyOR/}}
along with files in the original format, under the MIT license. 
We disclaim that the copyright to the content belongs to the original author 
John~E. Beasley.\footnote{\texttt{http://people.brunel.ac.uk/~mastjjb/jeb/orlib/legal.html}}

If one does not wish to download all the test problems,
we have also included one test instance ($A.4$ in the Library) in NumPy data format in the \texttt{BeasleyOR} directory,
on the repository webpage. 

% ==================================================================
\subsection{Test the code}
% ==================================================================

Once the code has been installed, the user can run the following test in a Python (3) command shell 
(assuming the test data has been downloaded in the \texttt{BeasleyOR} directory):

\begin{verbatim}
 > from SetCoverPy import setcover
 > a_matrix = np.load('./BeasleyOR/scpa4_matrix.npy')
 > cost = np.load('./BeasleyOR/scpa4_cost.npy')
 
 > g = setcover.SetCover(a_matrix, cost)
 > solution, time_used = g.SolveSCP()
\end{verbatim}

\noindent The variables and functions should be self-explanatory. 
The test should run without a problem and yield a (near-)optimal solution in less than 1 minute 
on a modern laptop with some average configuration (as of 2016).
Once the solver finishes, the attribute \texttt{g.s}, a binary 1D array with the same size
as the number of columns, gives the current solution of the minimum set, and 
\texttt{g.total\_cost} gives the value of the corresponding minimum cost.

% ==================================================================
\subsection{The test dataset of the extragalactic sources}
% ==================================================================

We also make the test dataset publicly available.\footnote{\texttt{http://www.pha.jhu.edu/\%7egz323/scp/Data/ExtragalacticTest/}}
The dataset includes fields that can be used to identify the objects, such as RA and DEC, 
and also stellar mass and star formation rate estimates from the MPA-JHU value-added catalog,
and also the optical spectra observed by SDSS but interpolated onto the same wavelength grid.

In addition, in SetCoverPy, we also include two routines for (quick) estimation of the weighted $\chi^2$
distance, \texttt{quick\_amplitude} and \texttt{quick\_totalleastsquares}. 
If you have two vectors $\bs{x}$ and $\bs{y}$ with errors $\bs{x}_{\rm err}$ and $\bs{y}_{\rm err}$, 
they perform a least squares fitting for the amplitude $a$ in $\bs{y} = a\bs{x}$ in an iterative manner.
For example:

\begin{verbatim}
 > from SetCoverPy import mathutils 
 > a, chi2 = mathutils.quick_amplitude(x, y, xerr, yerr)
\end{verbatim}
\noindent  or 
\begin{verbatim}
 > a, chi2 = mathutils.quick_totalleastsquares(x, y, xerr, yerr)

\end{verbatim}

\noindent The difference between the two functions is the second includes an optimization step 
with the \texttt{optimize} module in SciPy
and is considerably slower, though it provides (slightly) more accurate results. 
For most robust fitting we recommend the reader to use a Bayesian estimator \citep[\eg][]{hogg2010a, foreman2013a}.

\bibliographystyle{apj}
%\bibliography{archetype.bib}

\end{document}